\documentstyle[12pt]{article}
\def\bc{\begin{center}}
\def\ec{\end{center}}
\def\be{\begin{equation}}
\def\ee{\end{equation}}
\def\bea{\begin{eqnarray}}
\def\eea{\end{eqnarray}}

\def\simge{\ \lower-
1.2pt\vbox{\hbox{\rlap{$>$}\lower5pt
\vbox{\hbox{$\sim$}}}}\ }
\begin{document}
\pagestyle{empty} 
\vspace{-0.6in}
\begin{flushright}
CERN-TH/97-246 \\
%ROME 97/1175 \\
%TUM-HEP-279/97 
\end{flushright}
\vskip 2.0in
\centerline{\large {\bf{UNSTABLE 
SYSTEMS}}}
\centerline{\large {\bf {IN}}}
\centerline{\large {\bf {RELATIVISTIC 
QUANTUM FIELD THEORY}}}
\vskip 1.0cm
\centerline{L. Maiani$^1$, M. Testa$^{1,2}$}
\centerline{\small $^1$ Dipartimento di 
Fisica, Universit\`a di Roma ``La
Sapienza"}
\centerline{\small Sezione INFN di Roma}
\centerline{\small P.le A. Moro 2, 00185 
Roma, Italy}
\centerline{\small $^2$  Theory Division, 
CERN, 1211 Geneva 23,
Switzerland$^{\star}$.}
\vskip 1.0in
\abstract{We show how the state of an 
unstable particle can be defined in
terms of stable asymptotic states. This 
general definition is used to
discuss and to solve some old problems 
connected with the short-time and
large-time behaviour of the non-decay 
amplitude.}
\vskip 1.0in
\begin{flushleft} 
CERN-TH/97-246 \\
%ROME 97/1175 \\
%TUM-HEP-297/97 \\
September 1997
\end{flushleft}
\vfill
\noindent \underline{\hspace{2in}}\\
$^{\star}$ Address from Sept 1st, 1997 
to August 31st, 1998.
\eject
\pagestyle{empty}\clearpage
\setcounter{page}{1}
\pagestyle{plain}
\newpage 
\pagestyle{plain} \setcounter{page}{1}

\section{Introduction}

The definition of unstable states in quantum 
mechanics is notoriously
difficult\cite{sch},\cite{zero},\cite{zeroa}. The most 
usual approach is based on the
Breit-Wigner approximation of scattering 
amplitudes\cite{weinberg} in which the 
unstable
state is associated with a factorizable simple 
pole in the complex $s$-channel
variable, $s$ being the total momentum 
squared in
$s$-channel. This approach, however, is not 
suitable to describe the detailed
space-time behaviour of the unstable system. 
In particular it underscores the
deviations from simple exponential 
behaviour at very short and very large times,
which follow from general 
arguments\cite{green}.

The short-time behaviour of an unstable 
state weakly coupled to the final decay
channels has been investigated recently in 
Ref.\cite{bern}. The result has
been to underline an unexpected singularity 
in the short-time expansion of the
non-decay amplitude, which makes the 
deviations from exponential
behaviour in this region depend from the 
formation process of the unstable state
itself. The short-time singularity and the 
corresponding dependence
from the formation condition of the unstable 
state seems to eliminate the
so-called Zeno paradox (\cite{uno}-
\cite{otto}) at least for systems described
in terms of Relativistic Quantum Field Theory.

In this paper, we present a general, 
relativistically invariant, definition of
the unstable state which allows to elucidate 
the space-time
behaviour of the non-decay amplitude. The 
definition follows
closely the experimental procedure to 
measure the lifetime of weakly decaying
particles (i.e. the so-called impact parameter 
distribution). 

For weakly unstable particles, where 
perturbation theory can be used, we find
that the general quantum-mechanical 
expression of the decay rate per unit time
(the Fermi Golden Rule) has to be modified at 
short times in a well defined way.
In addition, we recover the results of the 
previous analysis\cite{bern}, with a
much clearer characterization of the non-
universal dependence upon the
wave-packet shape of the initial particles 
from which the unstable state was
formed.

The space-time behaviour of the non-decay 
amplitude is specified in terms of
$S$-matrix amplitudes. Adopting for the 
latter the Breit-Wigner form with a
constant width gives, for times larger than 
the formation time, the familiar
exponential behaviour.

Finally, on the basis of unitarity arguments, 
one can estimate the very large
time behaviour where again deviations from 
the exponential form are to be
expected, in the form of a power-law time 
dependence. It is worth stressing
that, in this case, the time  behaviour is again 
a non-universal one,  related
to the form of the initial state wave packets. 

Although not strictly related to the central 
arguments of the present paper, we
also give, in the last section, a simple and 
general derivation of the
so-called Wigner's delay relation\cite{wigner} 
for scattering systems,
because it fits very naturally within the 
general spatio-temporal description of
the scattering process used in this 
investigation.

\section{Constructing the Unstable State from 
Scratch}

We define as usual the $S$- and the $T$-
matrices, according to:
\begin{eqnarray}
& & S_{\alpha \beta }=\left\langle {{\alpha 
;out}} \mathrel{\left | {\vphantom
{{\alpha ;out} {\beta ;in}}} \right. \kern-
\nulldelimiterspace} {{\beta ;in}}
\right\rangle =\left\langle {\alpha ;in} 
\right|S\left| {\beta ;in}
\right\rangle =\left\langle {\alpha ;out} 
\right|S\left| {\beta ;out}
\right\rangle \nonumber\\
& & S=I+iT=\sum\nolimits_\delta  {\left| 
{\delta ;in} \right\rangle
}\left\langle{\delta ;out} \right| \nonumber\\
& & TT^+=T^+T=2 Im T\label{uno}
\end{eqnarray}
where $\alpha $, $\beta$ and $\delta$ 
denote a suitable set of free particle
quantum numbers, e.g. momenta and spin 
components. However, to describe the
space-time evolution of the scattering process 
it is necessary to introduce
wave-packtes for the initial states. We 
consider, therefore, a two-particle
state in the far past:
\begin{equation}
\left| {in} \right\rangle =\int 
{d^3p_1d^3p_2f(\underline p_1)g(\underline
p_2)\left| {\underline p_1,\underline p_2;in} 
\right\rangle }\label{due}
\end{equation}

We consider, for simplicity, equal mass, 
spinless, particles and we set
ourselves in the center of mass (c.o.m.) frame 
of reference. We are interested
in a situation in which the incoming particles 
can create a resonant state. We
therefore choose the wave packets so that 
$f(\underline p)$ is peaked around some
momentum ${\underline p}_{res}$ and
$g(\underline p)$ around momentum $- 
{\underline p}_{res}$ such that the c.o.m.
energy is about equal to the resonance mass 
$M$:
\begin{equation}
2\omega ({\underline p}_{res})\approx \sqrt 
s \approx M \label{tre}
\end{equation}

Wave packets are choosen so as to represent 
very distantly localized particles,
at some large, negative time $t= -T$, and to 
overlap around the origin of
coordinates, at time $t=0$. Neglecting long-
range forces, which can be dealt with
separately, wave packets evolve in time 
freely for all times before collision, up
to the last $10^{-23}$ seconds or so. We 
consider the state:
\begin{equation}
-iT\left| {in} \right\rangle =(I-S)\left| {in} 
\right\rangle =\left| {in}
\right\rangle -\left| {out} 
\right\rangle\equiv \left| R \right\rangle 
\label{quattro}
\end{equation}

The state $\left| {out} \right\rangle $, in 
Eq.[\ref{quattro}], represents a two
particle state in the distant future in the 
wave packets $f$ and $g$, such that
they were overlapping around the origin at 
$t=0$. Thus the state vector $\left| R
\right\rangle $ defined by Eq.[\ref{quattro}] 
is the initial state minus the
state where nothing happens, i.e. it 
represents the state of the products of the
collision  which has taken place around time 
$t=0$.

We consider next the amplitude:
\begin{equation} A(t)=\left\langle {in} 
\right|T^+e^{-iHt}T\left| {in}
\right\rangle\equiv \left\langle {{Rt}} 
\mathrel{\left | {\vphantom {{Rt} R}}
\right. \kern-\nulldelimiterspace} {R} 
\right\rangle\label{cinque}
\end{equation}

$A(t)$ represents the overlap of the collision 
state, when the collision has
taken place at time $t=0$, $\left| R 
\right\rangle $, with the collision state when 
the collision has taken
place at a later time $t$ (the state 
$\left\langle {in} \right|iT^+e^{-iHt}
\equiv \left\langle {Rt} \right|$). If the 
collision goes through the formation
of an unstable state, $A(t)$ is, apart from a 
normalization factor, the
amplitude for this state to have remained 
unchanged during time $t$, i.e. the
non-decay amplitude, so that:
\begin{eqnarray}
A_{non-decay}(t)={{A(t)} \over {A(0)}} 
\nonumber \\
P_{non-decay}(t)=\left| {{{A(t)} \over {A(0)}}} 
\right|^2 \label{sei}
\end{eqnarray}

Eq.[\ref{sei}] is our basic starting point. We 
write:
\begin{equation}
\left\langle {\underline p_3,\underline 
p_4;in} \right|T\left| {\underline
p_1,\underline p_2;in} \right\rangle =(2\pi 
)^4\delta
^{(4)}(p_3+p_4-p_1-p_2) T(s,t)\label{sette}
\end{equation}
where $s$ and $t$ are the usual Mandelstam 
variables. Using the unitarity
relation, Eq.[\ref{cinque}] can then be 
rewritten as:
\begin{eqnarray}
& & A(t)=\int 
{d^3p_1d^3p_2d^3p_3d^3p_4f^*(\underline 
p_3)g^*(\underline
p_4)f(\underline p_1)g(\underline p_2)}e^{-
i(\omega _{p_1}+\omega
_{p_2})t}\times\nonumber\\
& & \times(2\pi )^4\delta ^{(4)}(p_3+p_4-
p_1-p_2)2 ImT(s,t)=\label{otto}\\
& & =\left\langle {e^{-i(\omega 
_{p_1}+\omega
_{p_2})t}(2\pi)^4\delta^{(4)}(p_3+p_4-p_1-
p_2)2 Im T(s,t)} \right\rangle
\nonumber
\end{eqnarray}
where brackets indicate folding with wave 
packets. Eq.[\ref{otto}] can be
rewritten as:
\begin{equation}
A(t)=e^{-iMt}\int {dxe^{-ixt}F(x)}\label{nove}
\end{equation}
with $F(x)$ a positive definite function with a 
limited-from-below support
(corresponding to the combined support of 
$Im T$ and of the wave packets):
\begin{equation}
F(x)\equiv \left\langle {(2\pi )^4\delta 
^{(4)}(p_3+p_4-p_1-p_2)
\delta(\omega_{q_1}+\omega_{q_2}-M-x) 2 
Im T(s,t)}\right\rangle\label{dieci}
\end{equation}

We add a few comments.
\begin{itemize}
\item The definition of the non-decay 
probability,
Eqs.[\ref{cinque}],[\ref{sei}], is given in term 
of $S$-matrix elements, i.e.
it involves only the stable asymptotic states 
out of which the unstable state is
formed (and in which it will, eventually, 
decay). No assumption is made, in
particular, about the existence of a state 
representing the unstable particle
itself (at time
$t=0$), which is indeed a very questionable 
assumption, certainly not valid
beyond perturbation theory. If we had 
assumed that, we could have written:
\begin{equation}
A(t)=\left\langle R \right|e^{-iHt}\left| R 
\right\rangle =\sum\nolimits_n
{e^{-iE_nt}\left| {\left\langle {n} 
\mathrel{\left | {\vphantom {n R}} \right.
\kern-\nulldelimiterspace} {R} \right\rangle 
} \right|^2} \label{undici}
\end{equation}
The amplitude $A(t)$ in Eq.[\ref{cinque}], as 
shown by Eqs.[\ref{nove}] and
[\ref{dieci}], shares with the naive amplitude, 
Eq.[\ref{undici}], the property
of being the Fourier-transform of a positive 
definite function with a
limited-from-below support. This property, 
first stressed by
L.A. Khalfin\cite{uno} for the naive amplitude 
Eq.[\ref{undici}], gives rise to
the non-exponential behaviour af $A(t)$ at 
very large times, as discussed
in Section \ref{sec:nonexp}.
\item The above
definition of $A(t)$ makes sense only for 
times $t$ which are larger than either
the overlap time of the initial wave packets 
or the characteristic decay time of
the background processes. In turn, this 
implies that the definition is useful
only if the lifetime of the resonance is larger 
than either these characteristic
times.

Since:  
\begin{equation}
\left. {\Delta t} \right|_{overlap}={{\Delta x} 
\over v}\approx {1 \over {v\Delta p}}\approx 
{1
\over {\Delta E}}\approx {1 \over M}  
\label{dodici}
\end{equation}
the first condition gives:
\begin{equation}
\Gamma ={1 \over \tau }<<{1 \over {\left. 
{\Delta t}
\right|_{overlap}}}\approx M \label{tredici}
\end{equation}
Similarly, background processes (e.g. box-
diagram contributions to the
scattering process) decay in time with the 
only time-scale available, that is:
\begin{equation}
\left. {\Delta t} \right|_{backgnd}\approx {1 
\over {\sqrt s}}\approx {1 \over
M} \label{quattordici}
\end{equation}
so that, in conclusion, our definition is 
suitable for:
\begin{equation}
\Gamma ={1 \over \tau }<<M  \label{quindici}
\end{equation}
as intuitively expected.
\item If we transform from the c.o.m. to 
another frame
of reference, the time translation gives rise to 
a non vanishing
space translation. The amplitude $A(t)$ 
transforms into the probability amplitude
for the final particles to originate at a non-
vanishing distance from the
collision point. $P(t)$ corresponds, in this 
case, to the so-called impact
parameter distribution of the decay products 
of an unstable particle produced
with non-vanishing velocity.
\end{itemize}

\section{Perturbation Theory}

In this section we show how the previous 
general scheme works when
perturbation theory is reliable. In particular 
we will discuss the derivation of
the Fermi Golden Rule and its modifications 
at short times, related to the
singularities inherent to Relativistic Quantum 
Field Theory.

We shall study, for definiteness, a system 
consisting of an unstable scalar
particle
$\phi $, with mass $M$, decaying into two 
scalar particles $\psi _1$ and $\psi
_2$ with masses $m_1$ and $m_2$ 
respectively. We take a decay hamiltonian,
$H_I$, of the simple (super-renormalizable) 
form:
\begin{equation}
H_I=g\int {d^3x\,\phi \psi _1\psi 
_2}\label{sedici}
\end{equation}

The unstable particle $\phi $ contributes to 
the elastic scattering amplitude of the
two particles $\psi _1$ and $\psi _2$ as:
\begin{equation}
S_{fi}\equiv \left\langle {{p_3,p_4;out}} 
\mathrel{\left | {\vphantom
{{out\,p_3,p_4} {p_1,p_2\,in}}} \right. \kern-
\nulldelimiterspace}
{{p_1,p_2;in}}
\right\rangle =\delta _{fi}+{{i{\cal M}_{fi}} 
\over {\prod\limits_i {\sqrt {(2\pi
)^32\omega _i}}}}(2\pi )^4\delta 
^{(4)}(p_1+p_2-p_3-p_4)\label{diciassette}
\end{equation}
where:
\begin{equation}
i{\cal M}_{fi}=(ig)^2{i \over {P^2-
M^2+i\varepsilon }}-(ig)^2{i \over
{(P^2-M^2+i\varepsilon )}}i\Pi (P^2){i \over 
{(P^2-M^2+i\varepsilon
)}}\label{diciotto}
\end{equation}
and:
\begin{equation}
P=p_1+p_2=p_3+p_4\label{diciannove}
\end{equation}

\begin{equation}
i\Pi (P^2)\equiv \delta M^2+i^2\int 
{dx}\,\left\langle 0 \right|T(O(x)O(0))\left| 0
\right\rangle \exp (iPx)\label{venti}
\end{equation}

\begin{equation}
O(x)\equiv g\psi _1(x)\psi 
_2(x)\label{ventuno}
\end{equation}

The counterterm $\delta M^2$ in 
Eq.[\ref{venti}] is adjusted so that:
\begin{equation}
Re \Pi (M^2)=0\label{ventidue}
\end{equation}
and therefore $M$ is the renormalized mass 
of the resonant state.
The behaviour of $\Pi (P^2)$ around $M$ is 
then:
\begin{equation}
\Pi (P^2)\mathop \approx 
\limits_{P^2\approx M^2}(P^2-M^2) Re \Pi 
'(M^2)+i Im
\Pi (P^2)\label{ventitre}
\end{equation}
where $Re \Pi '(M^2)$ is an ultraviolet 
divergent parameter related to the wave
function renormalization of the unstable 
particle propagator, which is to be
removed by renormalization. It will be clear 
in a moment
that this term does not affect the non-decay 
amplitude of the unstable state. As
for $Im \Pi (P^2)$, it has the expression:
\begin{equation}
Im \Pi (P^2)={1 \over 2}\sum\limits_n {\left| 
{\left\langle 0
\right|O(0)\left| n \right\rangle } 
\right|^2(2\pi )^4\delta
^{(4)}(P-P_n)}\label{ventiquattro}
\end{equation}

The overlap $A(t)$ defined in 
Eq.[\ref{cinque}], then becomes:
\begin{eqnarray}
& & A(t)=\left\langle {{Rt}} \mathrel{\left | 
{\vphantom {{Rt} R}} \right.
\kern-\nulldelimiterspace} {R} \right\rangle 
=\int
{{{f^*(p_3)g^*(p_4)f(p_1)g(p_2)} \over {(2\pi 
)^6\sqrt {2\omega _{p_3}2\omega
_{p_4}2\omega _{p_1}2\omega 
_{p_2}}}}\,}\times\label{venticinque}\\
& & \times (2\pi )^4\delta ^{(4)}(p_1+p_2-
p_3-p_4)\exp [-i(\omega _{p_1}+\omega
_{p_2})t]\left\{ {-i{\cal M}_{fi}-(i{\cal 
M}_{fi})^*} \right\}\nonumber
\end{eqnarray}

We have:
\begin{equation}
-i{\cal M}_{fi}-(i{\cal M}_{fi})^*=(g)^2(2\pi 
)\delta (P^2-M^2)-2g^2 Im \left( {{{\Pi (P^2)}
\over {(P^2-M^2+i\varepsilon )^2}}} 
\right)\label{ventisei}
\end{equation}
and:
\begin{eqnarray}
& & Im \left( {{{\Pi (P^2)} \over {(P^2-
M^2+i\varepsilon )^2}}}
\right)=\nonumber\\
& & Im \Pi(P^2) Re {1 \over {(P^2-
M^2+i\varepsilon )^2}}+ Re \Pi
(P^2) Im {1 \over {(P^2-M^2+i\varepsilon 
)^2}}\label{ventisette}
\end{eqnarray}
so that:
\begin{equation}
Im \left( {{{\Pi (P^2)} \over {(P^2-
M^2+i\varepsilon )^2}}} \right)={{Im \Pi
(P^2)} \over {(P^2-M^2)^2}}-2\pi Re \Pi 
'(M^2)\delta (P^2-M^2)\label{ventotto}
\end{equation}

Eq.[\ref{venticinque}] then becomes:
\begin{eqnarray}
& & \left\langle {{Rt}} \mathrel{\left | 
{\vphantom {{RT} R}}
\right.\kern-\nulldelimiterspace} {R} 
\right\rangle =\int
{{{f^*(p_3)g^*(p_4)f(p_1)g(p_2)} \over {(2\pi 
)^6\sqrt {2\omega _{p_3}2\omega
_{p_4}2\omega _{p_1}2\omega 
_{p_2}}}}\,}\times\label{ventinove}\\
& & \times(2\pi )^4\delta ^{(4)}(p_1+p_2-
p_3-p_4)\exp [-i(\omega _{p_1}+\omega
_{p_2})t]\times\nonumber\\
& & \times\left\{ {g^2(1+2 Re \Pi '(M^2))2\pi 
\delta (P^2-M^2)-2g^2{{Im \Pi
(P^2)}
\over {(P^2-M^2)^2}}} \right\}\nonumber
\end{eqnarray}

For simplicity we choose to work in the 
narrow wave packet approximation. This
means that we parametrize the momenta as:
\begin{equation}
\underline p_i=\underline 
p_{res~i}+\underline k_i \label{trenta}
\end{equation}
and neglect terms quadratic in $\underline 
k_i$. 

Eq.[\ref{ventinove}] can therefore be 
rewritten as:
\begin{equation}
\left\langle {{Rt}} \mathrel{\left | 
{\vphantom {{Rt} R}} \right.
\kern-\nulldelimiterspace} {R} \right\rangle 
=\left\langle {{Rt}} \mathrel{\left
| {\vphantom {{Rt} R}} \right. \kern-
\nulldelimiterspace} {R} \right\rangle
_1+\left\langle {{Rt}} \mathrel{\left | 
{\vphantom {{Rt} R}} \right.
\kern-\nulldelimiterspace} {R} \right\rangle 
_2\label{trentuno}
\end{equation}
with:
\begin{eqnarray}
& & \left\langle {{Rt}} \mathrel{\left | 
{\vphantom {{Rt} R}} \right.
\kern-\nulldelimiterspace} {R} \right\rangle 
_1=2\pi \,g^2(1+2 Re \Pi
'(M^2))\exp [-iM\,t]\,\int 
{{{f^*(p_3)g^*(p_4)f(p_1)g(p_2)} \over {(2\pi
)^6\sqrt {2\omega _{p_3}2\omega 
_{p_4}2\omega _{p_1}2\omega
_{p_2}}}}\,}\times\nonumber\\
& & \times(2\pi )^4\delta ^{(3)}(\underline 
k_1+\underline
k_2-\underline k_3-\underline k_4)\delta 
[2M\underline v\cdot (\underline
k_1-\underline k_2)]\,\delta [\underline 
v\cdot (\underline k_3-\underline
k_4)]\label{trentadue}
\end{eqnarray}
and:
\begin{eqnarray}
& & \left\langle {{Rt}} \mathrel{\left | 
{\vphantom {{Rt} R}} \right.
\kern-\nulldelimiterspace} {R} \right\rangle 
_2=-2g^2\exp [-iM\,t]\,\int
{{{f^*(p_3)g^*(p_4)f(p_1)g(p_2)} \over {(2\pi 
)^6\sqrt {2\omega _{p_3}2\omega
_{p_4}2\omega _{p_1}2\omega 
_{p_2}}}}\,}\times\label{trentatre}\\
& & \times(2\pi )^4\delta ^{(4)}(p_1+p_2-
p_3-p_4)\exp -i[\underline v\cdot
(\underline k_1-\underline k_2)]t{{Im \Pi 
(P^2)} \over {[2M\underline v\cdot
(\underline k_1-\underline 
k_2)]^2}}\nonumber
\end{eqnarray}

Using Eq.[\ref{ventisette}], Eq.[\ref{trentatre}] 
becomes:
\begin{eqnarray}
& & \left\langle {{Rt}} \mathrel{\left | 
{\vphantom {{Rt} R}} \right.
\kern-\nulldelimiterspace} {R} \right\rangle 
_2=\nonumber\\
& & =-\,2\pi \,g^2\exp
[-iM\,t]\,\sum\limits_n {\left| {\left\langle 0 
\right|O(0)\left| n
\right\rangle } \right|^2(2\pi )^3\delta 
^{(3)}(\underline P_n)}{{\exp
-i(E_n-M)t} \over {4M^2(E_n-
M)^2}}\times\label{trentaquattro}\\
& & \times\int 
{{{f^*(p_3)g^*(p_4)f(p_1)g(p_2)} \over {(2\pi 
)^6\sqrt {2\omega
_{p_3}2\omega _{p_4}2\omega 
_{p_1}2\omega _{p_2}}}}\,}(2\pi )^4\delta
^{(4)}(p_1+p_2-p_3-p_4)\delta [M+\underline 
v\cdot (\underline k_1-\underline
k_2)-E_n]\nonumber
\end{eqnarray}

We can now define $\alpha$, $\alpha_0$, 
$\alpha_1$ and  $\beta_1$ through:
\begin{equation}
\left\langle {{Rt}} \mathrel{\left | 
{\vphantom {{Rt} R}} \right.
\kern-\nulldelimiterspace} {R} \right\rangle 
_1\equiv 2\pi g^2\alpha \,\exp
-iMt\equiv (2\pi g^2)(\alpha _0+\alpha 
_1)\,\exp -iMt\label{trentacinque}
\end{equation}
and
\begin{equation}
\left\langle {{Rt}} \mathrel{\left | 
{\vphantom {{Rt} R}} \right.
\kern-\nulldelimiterspace} {R} \right\rangle 
_2\equiv 2\pi g^2\beta _1(t)\,\exp
-iMt\label{trentasei}
\end{equation}
where $\alpha $ is independent of $t$, 
$\alpha _0$ is order $0$ in $g^2$, while
$\alpha _1$ and $\beta _1(t)$ are first order 
in $g^2$.

We finally have for the properly normalized 
non-decay probability defined in
Eq.[\ref{sei}]:
\begin{eqnarray}
& & P_{non-decay}(t)={{\left| {\left\langle 
{{Rt}} \mathrel{\left | {\vphantom
{{Rt} R}} \right. \kern-\nulldelimiterspace} 
{R} \right\rangle } \right|^2}
\over {\left| {\left\langle {{R0}} 
\mathrel{\left | {\vphantom {{R0} R}} \right.
\kern-\nulldelimiterspace} {R} \right\rangle 
} \right|^2}}={{\left|
{\left\langle {{Rt}} \mathrel{\left | 
{\vphantom {{Rt} R}} \right.
\kern-\nulldelimiterspace} {R} \right\rangle 
_1+\left\langle {{Rt}}
\mathrel{\left | {\vphantom {{Rt} R}} \right. 
\kern-\nulldelimiterspace} {R}
\right\rangle _2} \right|^2} \over {\left| 
{\left\langle {{R0}} \mathrel{\left |
{\vphantom {{R0} R}} \right. \kern-
\nulldelimiterspace} {R} \right\rangle
_1+\left\langle {{R0}} \mathrel{\left | 
{\vphantom {{R0} R}} \right.
\kern-\nulldelimiterspace} {R} \right\rangle 
_2} \right|^2}}\approx\nonumber\\
& & \approx {{1+2 Re [\,{{\alpha _1+\beta 
_1(t)} \over {\alpha _0}}]} \over {1+2
Re [\,{{\alpha _1+\beta _1(0)} \over {\alpha 
_0}}]}}\approx 1+2 Re [\,{{\beta
_1(t)-\beta _1(0)} \over {\alpha 
_0}}]\label{trentasette}
\end{eqnarray}

If, for simplicity, we take real momentum-
space wave functions, we get:
\begin{equation}
P_{non-decay}(t)=1-{2 \over 
M}\,\sum\limits_n {{{\left| {\left\langle 0
\right|O(0)\left| n \right\rangle } \right|^2} 
\over {(E_n-M)^2}}(2\pi )^3\delta
^{(3)}(\underline P_n)}\sin ^2[{{(E_n-M)} 
\over 2}t]\,H(E_n-M)\label{trentotto}
\end{equation}
where:
\begin{eqnarray}
& & H[E_n-M]\equiv\label{trentanove}\\
& & \equiv{{\int {{{f(p_3)g(p_4)f(p_1)g(p_2)} 
\over {(2\pi )^6\sqrt
{2\omega _{p_3}2\omega _{p_4}2\omega 
_{p_1}2\omega _{p_2}}}}\,}(2\pi )^4\delta
^{(4)}(p_1+p_2-p_3-p_4)\delta [M-
E_n+\underline v\cdot (\underline
k_1-\underline k_2)]} \over {\int 
{{{f(p_3)g(p_4)f(p_1)g(p_2)} \over {(2\pi
)^6\sqrt {2\omega _{p_3}2\omega 
_{p_4}2\omega _{p_1}2\omega 
_{p_2}}}}\,}(2\pi
)^4\delta ^{(4)}(p_1+p_2-p_3-p_4)\delta 
[\underline v\cdot (\underline
k_1-\underline k_2)]}}\nonumber
\end{eqnarray}
is a function which provides a cutoff in 
energy at small times $t$, when the
energy conservation is not yet active.
In fact, due to the finite spread in energy of 
the wave packets, $H(x)$ rapidly
vanishes for large $x$, while it goes to unity 
for vanishing $x$:
\begin{equation}
H[0]=1\label{quaranta}
\end{equation}

Eq.[\ref{trentotto}] reproduces the result first 
found in
Ref.\cite{bern}. The ''form-factor'',
$H(E_n-M)$, provides the cut-off to the sum 
over intermediate states which
is generally needed to cope with the singular 
behavior at very small times, as
further discussed in the next section. At 
larger times, $t$, on the other hand:
\begin{equation}
\left[ {2{{\sin [{{E_n-M} \over 2}t]} \over 
{E_n-M}}} \right]^2\to 2\pi
t\delta (E_n-M)\label{quarantuno}
\end{equation}
and Eq.[\ref{trentotto}], in virtue of 
Eq.[\ref{quaranta}], reduces to the usual
Golden Rule formula for the non-decay 
probability of an unstable particle:
\begin{equation}
P_{non-decay}(t)=1-{t \over 
{2M}}\,\sum\limits_n {\left| {\left\langle 0
\right|O(0)\left| n \right\rangle } 
\right|^2(2\pi )^4\,\delta ^{(3)}(\underline
P_n)}\delta [E_n-M]\label{quarantadue}
\end{equation}

\section{The Nature of Short-Time 
Singularities}

As discussed in Ref.\cite{bern}, the use of 
perturbation theory without
taking into account the formation time of the 
resonant state, would lead to an
expression for $P_{non-decay}(t)$ similar to 
the one given by
Eq.[\ref{trentotto}] with $H(E_n-M)\equiv 1$, 
i.e.:
\begin{equation}
P_{non-decay}(t)=1-{2 \over 
M}\,\sum\limits_n {{{\left| {\left\langle 0
\right|O(0)\left| n \right\rangle } \right|^2} 
\over {(E_n-M)^2}}(2\pi )^3\delta
^{(3)}(\underline P_n)}\sin ^2[{{(E_n-M)} 
\over
2}t]\label{quarantatreb}
\end{equation}

It was shown in Ref.\cite{bern} that the 
expansion of Eq.[\ref{quarantatreb}] in
powers of $t$ is usually marred by 
meaningless ultraviolet divergencies. In
this section we discuss in more detail the 
nature of these short-time
singularities. This discussion clarifies the 
nature of the assumptions needed
to derive the so called Zeno paradox 
(\cite{uno}-\cite{otto}). In fact the Zeno
paradox strongly depends on the quadratic 
short-time behaviour, usually inferred
from the (highly formal) argument:
\begin{eqnarray}
& & P_{non-decay}(t) = \mid{<P}\mid{e^{-
iHt}}\mid{P>}\mid{^{2}}=\nonumber\\
& & = 1- t^{2}(<P\mid{H^2} \mid{P>} - 
\mid{<P}\mid{H}\mid{P>}\mid{ ^2})
+...=\nonumber
\\ & & = 1 - \Delta E^{2}t^{2} + 
...\label{quarantacinqueb}
\end{eqnarray}

Expanding the explicit expression of $P_{non-
decay}(t)$ given by
Eq.[\ref{quarantatreb}] in powers of $t$, we 
have:
\begin{eqnarray}
& & P_{non-decay}(t)=1-{t^2 \over 
{2M}}\,\sum\limits_n {{\left| {\left\langle 0
\right|O(0)\left| n \right\rangle } 
\right|^2}(2\pi )^3\delta
^{(3)}(\underline P_n)}=\nonumber\\
& & =1-{{t^2} \over {2M}}\int {d\underline 
x\left\langle 0 \right|O(\underline
x,0)O(0)\left| 0 \right\rangle } 
\label{quarantaquattrob}
\end{eqnarray}

Eq.[\ref{quarantaquattrob}] shows that in this 
way we get, indeed, formally, a
$t^2$ short-time behaviour, as in 
Eq.[\ref{quarantacinqueb}]. The problem
is that the operator product appearing in 
Eq.[\ref{quarantaquattrob}] is not
integrable. In fact, apart from possible logs, 
we have from the Operator Product
Expansion:
\begin{equation}
O(\underline x,0)O(0)\mathop \approx 
\limits_{\underline x\approx 0}{1 \over
{\left| {\underline x} \right|^{2d_O}}}I+... 
\label{quarantacinquec}
\end{equation}
where $d_O$ denotes the dimension of the 
operator $O$. In the present example
$d_O=2$ and Eq.[\ref{quarantaquattrob}] 
suffers from a linear ultraviolet
divergence. The situation worsens for more 
singular (higher dimensional) decay
hamiltonians, as discussed in Ref.\cite{bern}.

These considerations clearly show that there 
is a short-time ultraviolet singularity which:
\begin{itemize}
\item makes $\Delta E^2$ divergent in 
Eq.[\ref{quarantacinqueb}]
\item is smeared, in the particle survival 
amplitude Eq.[\ref{trentotto}], over time 
scales
provided by the formation mechanism of the 
unstable state (the overlap time of the incident wave-packets). 
\end{itemize}

It should be clear that no form factors (in the 
case of decays involving
hadrons, as e.g. proton decay) can cure this 
divergence which is quite similar
to the one measured in deep-inelastic-
scattering, due to the singular product of
two currents.

In Eq.[\ref{trentotto}], we can interpret the 
presence of $H(E_n-M)$
as a cut-off on the energy of detected final 
decay products (a reasonable
requirement for a measuring apparatus), 
which, by the optical theorem,
produces an effective smearing on the 
particle survival amplitude.

We stress again the conclusions of 
Ref.\cite{bern} concerning the fallacy of
the finiteness of $\Delta E^2$ in 
Eq.[\ref{quarantacinqueb}], on which many of
the papers dealing with the Zeno paradox are 
based.

\section{The Exponential Decay}

Higher orders in perturbation theory give 
rise to higher powers of the time,
$t$, in Eq.[\ref{quarantadue}]. For times of 
the order of the lifetime $\tau$ we
need to sum up at least the leading terms, to 
get a meaningful result. 

The linear behaviour in time, in 
Eq.[\ref{quarantadue}], arises from the 
singular
behaviour in $(s-M^2)$ of the self-energy 
insertion. Thus, higher
powers in $t$ shall correspond to repeated 
self-energy insertions in the
lowest-order particle propagator. 

More precisely, the leading singular 
behaviour is given by insertion of the
imaginary part of the self-energy evaluated 
at $s=M^2$. Inserting the successive
terms of the expansion of $Im\Pi$ around 
$s=M^2$ would give rise to less
singular terms in $(s-M^2)$, i.e. to a 
subleading behaviour in time. The same
applies to the insertion of $Re\Pi$, since it 
vanishes at $s=M^2$  by mass
renormalization.

Using the above considerations, we find that 
the sum of the leading terms in time
is given by the Breit-Wigner  propagator:
\begin{equation}
A(t)=\left\langle {e^{-i(\omega 
_{p_1}+\omega _{p_2})t}(2\pi
)^4\delta ^{(4)}(p_3+p_4-p_1-p_2)2 Im 
[T(s,t)]_{BW}}
\right\rangle\label{quarantatrea}
\end{equation}

\begin{equation}
T_{BW}={{-1} \over {s-M^2+i\Gamma 
M}}\label{quarantaquattroa}
\end{equation}
so that (compare with Eq.[\ref{nove}]):
\begin{equation}
A(t)=e^{-iMt}\int {dxe^{-ixt}{{2M\Gamma } 
\over {4M^2(x^2+{{\Gamma ^2} \over
4})}}G(x)}\label{quarantacinquea}
\end{equation}
with $G(x)$ defined by:
\begin{equation}
G(x) \equiv \left\langle {(2\pi )^4\delta 
^{(4)}(p_3+p_4-p_1-p_2)\delta (\omega
_{p_1}+\omega _{p_2}-M-x)} 
\right\rangle\label{quarantaseia}
\end{equation}

In the narrow wave packet approximation, 
Eq.[\ref{trenta}], we have:
\begin{equation}
\omega _{p_1}+\omega _{p_2}\approx 
M+\underline v\cdot (\underline
k_1-\underline k_2)\label{quarantasettea}
\end{equation}
The region where $G(x) \approx constant 
\approx G(0)$ is thus limited
by:
\begin{equation}
x_{crit}\approx v\Delta p\approx {v \over 
{\Delta x}}\approx {1 \over {\left.
{\Delta t} \right|_{overlap}}}>>\Gamma 
\label{quarantottoa}
\end{equation}

Since the integral in 
Eq.[\ref{quarantacinquea}] is dominated by 
the region $x
\approx
\Gamma$, we may take $G(x)$ as a constant, 
to a good approximation. Extending,
furthermore, the integration range to $\pm 
\infty $, we find:
\begin{equation}
A(t)={{\pi G(0)} \over M}e^{-iMt} e^{-
{{\Gamma t} \over 2}} \label{quarantanovea}
\end{equation}
that is, in conclusion, a pure exponential 
behaviour for the non-decay
probability.

\section{Very Large Time 
Behaviour}\label{sec:nonexp}

The exponential behaviour displayed in 
Eq.[\ref{quarantanovea}] is the result of
an approximation\cite{sch},\cite{green}. In fact we 
kow from the Riemann-Lebesgue lemma that 
the
asymptotic behaviour of a Fourier transform 
integral, as the one appearing in
Eq.[\ref{nove}], is determined by the points 
where $F(x)$ or some of its
derivatives are singular. If $F(x)$ where 
continuous together with all its
derivatives, the non-decay amplitude 
$A_{non-decay}(t)$ whould vanish faster
than any power of ${1 \over t}$. The 
presence of any singularity makes
$A_{non-decay}(t)$ vanish not faster than 
some power of ${1 \over t}$.
From the expression of $F(x)$, Eq.[\ref{dieci}], 
it is clear that
singularities are necessarily present which 
originate from two possible
sources:
\begin{enumerate}
\item  singularities due to the unitarity of the 
$S$-matrix;
\item  singularities of the wave function of 
the initial state.
\end{enumerate}

As for 1), unitarity requires that a resonance 
pole be located on
the second sheet of the complex energy 
plane, thus implying the existence of at
least one branch singularity of the amplitude 
$T(s,t)$ in Eq.[\ref{dieci}].

The presence of this singularity will, 
however, be ineffective for the large
$t$ behaviour of $A_{non-decay}(t)$, unless 
the resonance location is very close
to threshold. In fact, as can be seen from 
Eq.[\ref{dieci}], its
contribution is, in general, depressed by the 
narrowness of the wave packet of
the initial state.

On the second point very little can be said. 
The possible singularities due
to the initial wave function have to do with 
the details of the experimental
preparation of the resonant state and are 
certainly not even under direct control of
the experimentalist.

Whatever their origin, the presence of these 
singularities in $F(x)$ make
$A_{non-decay}(t)$ behave asymptotically in 
a power-like way, rather than
exponentially. It must be remarked, however, 
that nothing general can be said on
the onset time of this power-like behaviour 
because it depends on the detailed
structure of the wave function of the initial 
state.

To get a crude idea of the times where the 
power law behaviour takes over, we may 
replace
Eq.[\ref{quarantaquattroa}] with a form in 
which the two particle cut is considered:
\begin{equation}
T={{-1}\over {s-M^2+iM\Pi(s)}} \label{one}
\end{equation}
where, to account for threshold behavour:
\begin{equation}
\Pi(s)=\Theta (s-
4m^2)[{{k(s)}\over{k(M)}}]^{2l+1}\Gamma
\label{two}
\end{equation}
and:
\begin{equation}
4k(s)^2=s-4m^2\label{three}
\end{equation}
but otherwise we neglect any energy 
dependence of the matrix element. With 
these positions, Eq.[\ref{quarantacinquea}] 
becomes:
\begin{equation}
A(t)=exp(-2imt)\int_0 {dx exp(-
ixt)[2M\Pi(s)][(s-M^2)^2+M^2\Pi(s)^2]^{-
1}G(x)}\label{four}
\end{equation}
with
\begin{equation}
\sqrt (s)\equiv x+2m\label{five}
\end{equation}

Neglecting further any variation of the 
function $G(x)$, which describes the energy 
spread of the initial wave packets, the 
asymptotic behaviour of $A(t)$, as $t \to 
\infty$, is easily estimated to be:
\begin{eqnarray}
A(t)_{asympt}=C[{{4m\Gamma} \over{ 
k(M}})]^{l+{5 \over 2}}(\Gamma t)^{-l-{3 
\over 2}}\label{six}
\end{eqnarray}
with:
\begin{equation}
C=({1\over{128}})({M \over m^2}) G(0) \int_0 
{du exp(-iu) u^{l+{1 \over 2}}}\label{seven}
\end{equation}
Comparing with the exponential law, 
Eq.[\ref{quarantanovea}], we see that the 
power law gets in for a critical time , 
$t_{crit}$, which is very large indeed. 
Numerically:
\begin{equation}
\Gamma t_{crit} \approx 2 (l+{5 \over 2})\{ 
25 + ln \{ [{{k(M)} \over {0.7 MeV}}][{{\tau} 
\over {10^{-10}}sec}] \} \}\label{eight}
\end{equation}

This result amply justifies the fact that no 
deviations from the exponential behaviour 
have been observed, until now, for long-lived 
systems\cite{nove}.

\section{Wigner's Delay}

Within the formalism described in the 
present paper, we can readily recover, in
a very general way the result, due to 
Wigner\cite{wigner} and usually derived
within potential scattering theory, which says 
that the derivative of scattering phase shift
$\delta _l(E)$ is related to the delay time 
induced by the interaction through:
\begin{equation}
T=2{{d\delta _l(E)} \over {dE}} 
\label{quarantadueb}
\end{equation}

In order to prove Eq.[\ref{quarantadueb}] let 
us consider the center of mass
scattering of two spinless particles in a state 
of given orbital angular
momentum $l$, under the inelastic threshold. 
The incoming state (a wave
packet normalized in a finite volume $V$) is:
\begin{equation}
\left| {f;in} \right\rangle=\sqrt {{{(2\pi )^3} 
\over V}}\int
{dE\,f(E)\left| {E,l,m;in}
\right\rangle}\label{quarantatre}
\end{equation}
where $\left| {E,l,m;in} \right\rangle$ denotes 
the incoming state
of two particles of given total energy $E$ and 
given angular momentum $l$, $m$,
normalized as:
\begin{equation}
\left\langle {{E',l',m';in}} \mathrel{\left | 
{\vphantom {{E',l',m';in}
{E,l,m}}} \right. \kern-\nulldelimiterspace} 
{{E,l,m;in}} \right\rangle={V
\over {(2\pi )^3}}\delta _{ll'}\delta 
_{mm'}\delta (E-E')\label{quarantaquattro}
\end{equation}
and
\begin{equation}
\int {dE\left| {f(E)}\right|^2}=1
\end{equation}

The amplitude to find, after the interaction, 
the system in a state which
corresponds to the free propagation of the 
initial state is:
\begin{eqnarray}
& & A=\left\langle {f;out} \right| {f;in} 
\rangle =\nonumber\\
& & ={{(2\pi )^3}\over V}\int {dE'dEf^*(E')f(E) 
\left\langle
{{E',l',m';out}}\mathrel{\left | {\vphantom 
{{E',l',m'} {E,l,m}}} \right.
\kern-\nulldelimiterspace} {{E,l,m;in}} 
\right\rangle}=\nonumber\\
& & =\int {dE\left| {f(E)}\right|^2\exp 2i\delta 
_l(E)}\label{quarantacinque}
\end{eqnarray}

If the system were non interacting, $\delta _l 
(E)=0$, and we would have $A=1$.

In the case of a narrow wave packet, in 
which $f(E)$ is strongly peaked around a
given energy $E_0$, we have:
\begin{equation}
A\approx \exp [2i\delta _l(E_0)-2i\delta 
'_l(E_0)E_0]\,\int {dE\left| {f(E)}
\right|^2\exp [2i\delta 
'_l(E_0)E]}\label{quarantasei}
\end{equation}

We can now ask what is the probability to 
find the system, after the
interaction, in a state which is the free 
propagation of the initial state
delayed by a time $T$. Such a time-
translated state is given by:
\begin{eqnarray}
& & \left| {f,T;out} \right\rangle=\exp 
iHT\sqrt {{{(2\pi )^3} \over V}}\int
{dE\,f(E)\left| {E,l,m;out} 
\right\rangle}=\nonumber\\
& & =\sqrt {{{(2\pi )^3} \over V}}\int
{dE\,f(E)\exp [iET]\,\left| {E,l,m;out} 
\right\rangle}\label{quarantasette}
\end{eqnarray}
where, as usual, $H$ denotes the full 
hamiltonian of the interacting system.

We have, in this case:
\begin{eqnarray}
& & A(T)=\left\langle {{f,T;out}} 
\mathrel{\left | {\vphantom {{f,T;out} {f;in}}}
\right. \kern-\nulldelimiterspace} {{f;in}} 
\right\rangle\approx\nonumber\\
& & \approx \exp [2i\delta _l(E_0)-2i\delta 
'_l(E_0)E_0]\,\int {dE\left|
{f(E)} \right|^2\exp iE[2\delta '_l(E_0)-
T]}\label{quarantotto}
\end{eqnarray}

From Eq.[\ref{quarantotto}] it follows that for 
$T=2\delta '_l(E_0)$, the
probability becomes $1$:
\begin{equation}
\left| {A(T=2\delta '_l(E_0))} 
\right|^2=1\label{quarantanove}
\end{equation}
which proves the Wigner delay relation, 
Eq.[\ref{quarantadueb}].

\end{document}